# RED-based Scheduler on Chip for Mixed-Criticality Real-Time Systems

Lukáš Kohútka, Lukáš Nagy, Viera Stopjaková
Institute of Electronics and Photonics
Slovak University of Technology in Bratislava
Bratislava, Slovakia
lukas.kohutka@stuba.sk

*Abstract*—**Real-time embedded systems that combine processes of various criticalities (i.e. mixed-criticality real-time systems) represent an emerging research that faces many issues. This paper describes a new ASIC design of a coprocessor that realizes process scheduling for mixed-criticality real-time systems. The solution proposed in this paper uses Robust Earliest Deadline (RED) algorithm. Due to the on-chip implementation of the scheduler, all scheduler operations always take two clock cycles to execute. The proposed solution was verified by simulations that applied millions of random inputs. Chip area costs are evaluated by synthesis into ASIC using 28 nm TSMC technology. The proposed RED-based scheduler is compared with an existing EDF-based scheduler that supports hard real-time processes only. Even though the RED-based scheduler costs more chip area, it can handle any combinations of process criticalities, variations of process execution times and deadlines, achieves higher CPU utilization and can be used for scheduling of non-real-time, soft real-time and hard real-time processes combined within one system.**

*Keywords—process scheduling, mixed-criticality, ASIC, RED.*

## I. INTRODUCTION

Mixed-criticality real-time systems are subject to very active research, as demonstrated in the review from Burns and Davis [1]. It is a global trend in microelectronics that results from increasing integration of more transistors on chip within to safety-critical applications. Certification of safety-critical systems usually expects process isolation to separate critical processes from non-critical processes using a separated hardware. However, process isolation forces systems to be under-utilized and to use pessimistic analysis of worst-case execution time for each process. On the other hand, average execution time is usually significantly smaller than the worst-case time [1-3].

Mixed-criticality real-time systems can support any combinations of critical and non-critical processes without using of process isolation. No process isolation is used in order to increase CPU utilization, achieved by execution of non-critical or low-priority processes (also known as best-effort processes) within the slack time that is available whenever a high-priority or safety-critical process is finished sooner than it is predicted according to the worst-case execution time analysis. This situation statistically keeps happening very often. The main issue is that best-effort processes and safety-critical processes have typically conflicting requirements [4-7].

## II. RELATED WORK

Earliest Deadline First (EDF) is one of the most popular scheduling algorithms for real-time systems, which is guaranteed to always set-up the best possible schedule in systems that use only hard RT processes [8, 9]. EDF algorithm keeps all processes sorted based on deadlines of these processes. Thus, the process with the earliest deadline (i.e. lowest deadline value) is selected for execution. The processes that are ready for execution are stored in a structure called "ready queue". Process queues can be implemented by various sorting architectures, such as shift registers architecture [10] or systolic array architecture [11].

In our previous work published in [12-15], we already presented novel RT process schedulers implemented in a form of coprocessors. We used a modified EDF algorithm with support to kill processes according to their ID. Beside our research, there are other solutions presented too. One solution employs EDF algorithm with the maximum number of processes being 64 [16], and the other approach uses priorities instead of deadlines, which is less efficient for real-time systems in terms of CPU utilization [17]. There are other solutions based on priorities or static scheduling [18-22] as well. However, these solutions are suitable for systems having hard RT processes only.

Because different types of processes are combined together in mixed-criticality real-time systems, a robust scheduler that can schedule all types of processes is required. Such a scheduler can be implemented on chip by using Robust Earliest Deadline (RED) algorithm, which can handle all types of processes and their combinations [1].

The RED algorithm can be viewed as an extension of the EDF algorithm. Both these algorithms are ordering processes according to process deadlines, as displayed in Fig. 1. EDF accepts all requests to add and schedule a newly created process, but RED also checks whether any of the scheduled processes can miss a deadline and if yes, then the most suitable process is temporarily rejected and moved to Reject queue. Thanks to this, it is guaranteed that every hard RT process always meets its deadline and as many soft RT processes as possible meet their deadlines as well. If one of the ready processes is finished sooner than the worst-case timing analysis predicts, and there is enough free time to reclaim and finish any of the rejected processes on time, then RED applies its reclaiming policy for reclaiming one of the rejected processes back from Reject Queue to the Ready queue [1, 2].

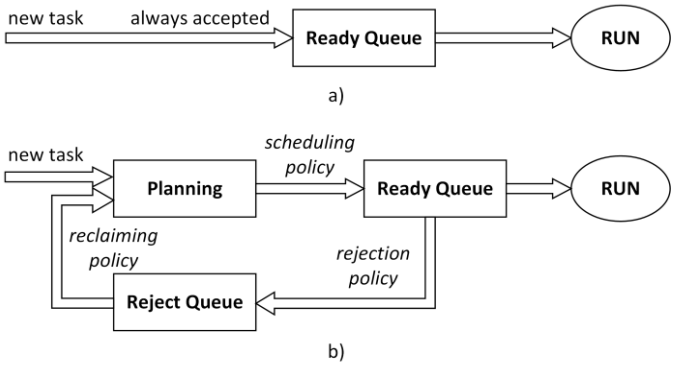

Fig. 1. EDF (a) and RED (b) scheduling algorithms [1].

### III. PROPOSED RED-BASED SCHEDULER

*1. Top module*

The proposed scheduler has a top module, which represents a coprocessor unit, as shown in Fig. 2, and it is composed of 3 sub-modules: Ready Queue, Reject Queue and Control Unit. The top module has only one input (except clock and reset, of course), called INSTR. The INSTR input represents one coprocessor instruction provided from CPU. There are 2 types of instructions supported by this coprocessor: *process insertion* and *process kill*. The top module has only 1 output, called PROCESS_TO_RUN, which represents the process that was selected by the scheduler for execution in CPU.

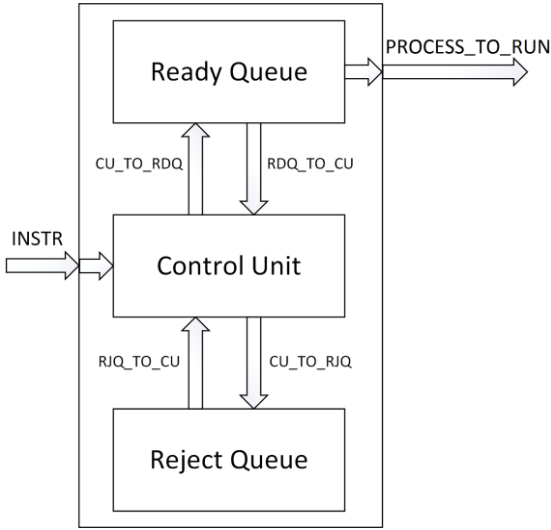

Fig. 2. RED scheduler top module.

*2. Ready Queue*

Ready Queue is a subcomponent that contains all ready processes (i.e. processes that are ready for execution), including the currently running process. The process with the lowest deadline is selected for execution. This subcomponent uses an existing priority-queue architecture that is called Shift Registers. Shift Registers consist of process cells, each containing one process comparator, control logic and a register to remember one process. Each process cell can exchange processes with adjacent cells. All process cells get the same instruction simultaneously from one common bus, driven from the input of Ready queue.

The Ready Queue subcomponent is extended with so-called "Overload Analysis" feature by applying these changes to Shift Registers:

1. Add a register for storing execution time in every process cell.
2. Add combination logic to detect overloads in every process cell.
3. Add combination logic to combine overload outputs from all process cells to create one single bit (overload).
4. Add combination logic to choose a victim in case of overload state (i.e. process rejection feature).

Step 1 – a new register (called execution time register) is added to each process cell within the Ready queue. This register represents the remaining worst-case execution time for the corresponding process cell plus the sum of the remaining worst-case execution times for all processes scheduled to be executed sooner than the actual process (i.e. all processes to the right in the queue). When a new process is inserted to one process cell, the value to be stored in execution time register is calculated by adding the worst-case execution time (WCET) of the new process and the value given by the execution time register from the process cell to the right from the actual process cell, where the new process is going to be inserted. In addition to this, the WCET value is also added to the actual values in the execution time registers in all process cells to the left (i.e. for all processes with higher deadline values). Whenever a process is supposed to be removed from the queue, all processes located in the process cells to the left from the process to be removed have to decrease their values in their execution time registers by the current execution time of the process that is removed.

Step 2 – a combinational logic is added to every process cell, which is responsible for calculation of overload bit for the given process cell. This logic is testing whether WCET of the new process and the value stored in the execution time register of the corresponding process cell are together higher than deadline of this process. If it is higher, then execution of the new process sooner than the execution of the process stored in the current process cell may cause that the deadline of the current process is not met. Such a situation is being referred as "overload" and an overload bit of the process cell is '1'.

Step 3 – one OR gate outside of the process cells but still inside the Ready queue subcomponent is added. This OR gate receives as inputs all the individual overload bits generated by the process cells (i.e. step 2). Therefore, if one or more overload bits are '1', then the OR gate output is '1'. Thus, there is an "overload" in the system in such a case. This overload bit is given to the Control Unit subcomponent, using RDQ_TO_CU interface.

Step 4 – a combinational logic is added, which performs a decision to choose one "victim process" from all processes located in Ready Queue. The victim process is removed from the Ready Queue and inserted to Reject Queue.

*3. Control Unit*

This subcomponent controls the other two subcomponents, i.e. Reject Queue and Ready Queue. Instructions from CPU are given to Control Unit, which further forwards instructions to Ready Queue. If no instruction is obtained from CPU, Control Unit can automatically move processes between Reject Queue and Ready Queue in order to maximize the portion of processes that are in Ready Queue but also to avoid system overloads by moving processes to Reject Queue.

Whenever the system is overloaded, the Control Unit gives an order to move one process (i.e. the victim) from Ready Queue to Reject Queue. This is performed by removing the process in Ready Queue and inserting the same process to Reject Queue. The selection of victim process is done according to priority values and position of the processes within the Ready Queue. The process that has the lowest priority value within those processes that are scheduled before the first process reporting the overload bit is selected as a victim process.

On the other hand, when the system is not overloaded and the Reject Queue contains at least one process, the Control Unit is trying to reclaim one process among the rejected ones by moving the process from Reject Queue back to the Ready Queue, if possible. This process is selected by the Reject Queue itself. Of course, the overload state needs to be evaluated after reclaiming the process again. If overload was detected again, then Control Unit must redo the reclaiming by moving the selected process from Ready Queue back to Reject Queue. However, if overload was not detected, then the reclaiming was successful and the reclaimed process may remain in Ready Queue.

There are four criticality levels available in the proposed process scheduler. These criticality levels are considered during the abovementioned process rejection and process reclaiming. These criticalities are encoded into two bits, using the following encoding:

- "00" – non-RT process or soft RT process of low priority
- "01" – soft RT process of medium priority
- "10" – soft RT process of high priority
- "11" – hard RT process (i.e. safety-critical)

*4. Reject Queue*

Reject Queue is a subcomponent that functions as a queue containing all rejected processes inside. The rejected processes are stored in Reject Queue so that they can be reclaimed (i.e. moved back to Ready Queue) without causing system to be overloaded. The rejected processes are being ordered in such a way that the process that has the highest priority and the latest deadline is available via output of the Reject Queue. This queue is implemented by Shift Registers, just like the Ready Queue. However, the Shift Registers architecture is configured as MAX Queue, not MIN Queue. The Shift Registers are ordering processes according to a combination of process deadline and process priority. The primary ordering criterion is process priority. Then, those processes that have the same priority are subsequently ordered according to process deadlines, i.e. deadlines are the secondary ordering criterion.

IV. VERIFICATION AND SYNTHESIS RESULTS

The functionality of the proposed solution in a form of a coprocessor unit has been verified by applying more than one million of simulation iterations. Every such iteration contained more than 500 coprocessor instruction that were generated pseudo-randomly. 50% of these instructions insert a new process, while the other 50% are killing a process.

The proposed RED scheduler and original EDF scheduler were synthesized in TSMC 28nm high performance mobile (HPM) technology using Cadence Genus tool. The proposed scheduler used clock frequency of 500 MHz. The synthesis used a power supply equal to 0.9 V. The chip areas of the EDF scheduler and RED scheduler are demonstrated in Table I. Both schedulers use the lowest-possible number of bits for process ID. The total power consumption of both schedulers using the same parameters is presented in Table II.

TABLE I. CHIP AREA OF EDF AND RED SCHEDULERS

| Number of Processes | EDF Chip Area [µm²] | RED Chip Area [µm²] | Overhead |
| --- | --- | --- | --- |
| 8 | 1702 | 13214 | +676,38% |
| 16 | 3637 | 28531 | +684,47% |
| 24 | 5979 | 50218 | +739,91% |
| 32 | 9085 | 69389 | +663,78% |
| 40 | 13307 | 92389 | +594,29% |
| 48 | 19387 | 108522 | +459,77% |
| 56 | 26787 | 130421 | +386,88% |
| 64 | 33340 | 146146 | +338,35% |

TABLE II. POWER CONSUMPTION OF EDF AND RED SCHEDULERS

| Number of Processes | EDF Power Consumption [µW] | RED Power Consumption [µW] | Overhead |
| --- | --- | --- | --- |
| 8 | 2044 | 12121 | +493,00% |
| 16 | 3809 | 25913 | +580,31% |
| 24 | 6356 | 43373 | +582,39% |
| 32 | 9881 | 66559 | +573,61% |
| 40 | 14711 | 77434 | +426,37% |
| 48 | 20693 | 93948 | +354,01% |
| 56 | 28263 | 115748 | +309,54% |
| 64 | 36535 | 124162 | +239,84% |

The overall comparison of the EDF-based and the novel RED-based schedulers is presented in Table III.

TABLE III. EDF AND RED SCHEDULERS OVERALL COMPARISON

| Criterion | Selected Scheduler | |
|---|---|---|
| | *EDF* | *RED* |
| Chip area | A | (4.38 – 8.40) x A |
| Power | P | (3.40 – 5.82) x P |
| Execution time | 2 clock cycles | 2 clock cycles |
| Hard RT processes | yes | yes |
| Soft RT processes | no | yes |
| Non-RT processes | yes | yes |
| Mixed-criticality supported | no | yes |

The proposed RED scheduler has higher chip area than the EDF scheduler. However, the RED scheduler supports mixed-criticality and all process types.

## V. CONCLUSION

We proposed a new ASIC architecture and implementation of a scheduler that can be used for complex mixed-criticality real-time systems. The research presented in this paper was focused on scalable reusing and extension of existing EDF schedulers based on Shift Registers architecture. The proposed scheduler can perform the scheduling in two clock cycles and uses an existing RED algorithm by extending EDF algorithm with process rejection and process reclaiming functionality. All real-time processes are categorized according to four criticality levels among the real-time processes and non-real-time processes are divided to 1024 priority levels. Thus, there are 1028 levels of criticality/priority in total. This level is set when processes are created. The RED scheduler is able to efficiently schedule any combination of hard RT, soft RT and non-RT processes. Therefore, the proposed RED scheduler is much more suitable for mixed-criticality RT systems.

The proposed coprocessor has still acceptable chip area that scales linearly with growing number of processes to be handled by the scheduler. In comparison to the existing software-based schedulers, the ASIC implementation is able to perform all operations with constant and significantly lower latency and with constant and much higher throughput, causing higher determinism of a system that would use the proposed scheduler. The proposed scheduler can be integrated to existing software-based operating systems as well.

## ACKNOWLEDGMENT

This work was supported in part by the Ministry of Education, Science, Research and Sport of the Slovak Republic under grant VEGA 1/0905/17, and ECSEL JU under project PROGRESSUS (876868).